# RLT4Rec: Reinforcement Learning Transformer for User Cold Start and Item Recommendation

Dilina Chandika Rajapakse
rajapakd@tcd.ie
Trinity College Dublin
Ireland

Douglas Leith
doug.leith@tcd.ie
Trinity College Dublin
Ireland

## ABSTRACT

We introduce a new sequential transformer reinforcement learning architecture RLT4Rec and demonstrate that it achieves excellent performance in a range of item recommendation tasks. RLT4Rec uses a relatively simple transformer architecture that takes as input the user's (item,rating) history and outputs the next item to present to the user. Unlike existing RL approaches, there is no need to input a state observation or estimate. RLT4Rec handles new users and established users within the same consistent framework and automatically balances the "exploration" needed to discover the preferences of a new user with the "exploitation" that is more appropriate for established users. Training of RLT4Rec is robust and fast and is insensitive to the choice of training data, learning to generate "good" personalised sequences that the user tends to rate highly even when trained on "bad" data.

## 1 INTRODUCTION

In this paper we revisit the use of Reinforcement Learning in recommender systems and introduce RLT4Rec. RLT4Rec adapts the sequential transformer neural network architecture with self attention, that has proven to be so effective in natural language tasks, and uses it to generate high quality item recommendations in a simple and robust manner.

The intuition for using a sequential transformer approach is that transformers have already demonstrated their ability to generate "good" token sequences (for a suitable definition of "good" e.g. in large language models it might be "grammatical human-like sentences"). Recommender systems also aim to generate "good" token sequences in the sense that the tokens are now items and we want the items to attract high user ratings. It is therefore natural to consider applying transformer models to directly generate good item sequences.

The generation of good item sequences can also be formulated as a Reinforcement Learning (RL) task. The items correspond to the actions in RL jargon, and the information gained about a user from observations of their past (item,rating) pairs is the RL state[1]. The RL reward to be maximised is the sum of the item ratings.

However, while there has been a good deal on interest in formulating recommendation as an RL task, to date RL techniques have not been widely adopted in practice. In part this is because of the difficulty of applying RL classical methods (such as DQN, MCTS, Actor-critic methods) for solving the recommendation task. For example, in recommender tasks the correct formulation of the user information state is often unclear, direct state observations are lacking and although one can try to estimate the state this adds to the complexity of the RL solution. Similarly, classical RL techniques are online methods i.e. in a recommender system they learn by interacting with live users. But such online training often provides a poor user experience, making it impractical. This has spurred consideration of offline RL where training uses previously collected data, but offline RL is known to suffer from thorny bias and convergence issues [15, 22].

By directly tackling the item recommendation task, our sequential transformer approach bypasses the intermediate MDP formulation step in RL and avoids many of the difficulties experienced with application of classical RL techniques to recommender systems. We are not the first to notice the potential for using sequential transformer methods for RL tasks generally, see for example the decision transformer [5] and trajectory transformer [11], but there has been surprisingly little work on their application to RL for recommender systems[2]. Recently, [23, 24] have applied decision transformer ideas to RL for recommender systems. These papers focus on offline RL training, which they recast as a supervised learning problem. However, they still require an explicit state formulation and state observations (using various LSTMs for state estimation), and in [23] offline training remains complex. They also do not evaluate per user recommendation performance vs the number of items rated by a user, a key aspect in recommender systems generally and in RL in particular. They also do not consider constraints e.g. that the same item should not be recommended excessively to a user (they allow the same item to be recommended repeatedly, without limit, which may skew the performance evaluation). RLT4Rec avoids all of these issues, while also making use of a simpler neural network architecture.

In summary, the main contributions of the paper are as follows:

(1) We introduce a new sequential transformer architecture RLT4Rec and demonstrate that it achieves state of the art performance in a range of item recommendation tasks. In particular, RLT4Rec consistently improves upon the performance of PRL [24] for established users and MCTS [19] for new users.

---

[1]This can be augmented with context information such as the user location, demographics etc.

[2]The bulk of the recommender system literature on transformers applies these to sequence-based recommendation rather than to RL, e.g. see Bert4Rec [21], SASRec [12]. In this literature the aim is essentially to mimic item sequences from the training data rather than to solve an RL task.

(2) RLT4Rec takes as input a sequence of user (item,rating) pairs and outputs the next item to present to the user i.e. it does not require input of a state observation or estimate. RLT4Rec is to the best of our knowledge the first RL approach that obviates this need for a state input.

(3) RLT4Rec handles new users and established users within the same consistent framework, it does not require a separate cold-start mode. New users just input a shorter (item,rating) history (or an empty history for completely new users) to RLT4Rec. RLT4Rec nevertheless achieves state of the art cold-start performance. That is, it automatically and efficiently balances the "exploration" needed to discover the preferences of a new user with the "exploitation" that is more appropriate for established users.

(4) Training of RLT4Rec is robust and fast. It can be trained using random sequences of items yet nevertheless learns to generate "good" personalised sequences that the user tends to rate highly. While offline RL learning is known to be difficult, our results show that this may be much less troublesome for RLT4Rec.

## 2 RELATED WORK

*Use of Transformers in RL Generally* Classical RL approaches formulate the online learning task as a Markov-decision process. The aim is to select a sequence of actions that maximises a cumulative reward in the face of a changing environment (changes may be random in nature as well as in response to previous actions taken). This can be solved using a range of techniques, e.g. Deep Q-Learning Networks (DQN). However, recently there has been much interest in alternatives approaches that make use of transformer neural networks. Probably the earliest works proposing a transformer approach to RL are Trajectory Transformer [11] and Decision Transformer (DT) [5], which have since led to a rapidly growing literature on transformer-based RL (eg: [16, 25, 26]). These transformer RL methods apply a sequence of (state, reward, action) triples as input to a transformer, each triple corresponding to a particular time instant. The last elements in the sequence are a (state, reward) pair and the task of the transformer is to predict the next action i.e. to complete the final (state, reward) pair to generate a predicted (state, reward, action) triple. In all of these approaches the state is assumed to be observed and so available as an input to the transformer.

*Transformers in RL for Recommender Systems* In recent years there has been much interest in the use of RL in recommender systems, see [1] for a survey. In RL the state of a user is an information state and consists of the history of items presented to a user and their rating (or other action e.g, a click or purchase). In the recommender system RL literature this sequence is mapped to a dense vector e.g. by use of an LSTM. See [10] for a recent survey on recommender system RL state estimation. The majority of existing work has focused on classical RL methods such as DQN and there has been much less work studying the use of transformers. In [24] a single (state, reward) pair is input to a transformer which then generates an action i.e. the next item to present to the user. The state is estimated using a variety of LSTM's. The main focus of the paper is on offline training, namely by reformulating the usual RL online learning task as supervised training of the transformer. In [23] a sequence of (state, reward, item) triples is input to a transformer.

The state is estimated using an LSTM. Again, the main focus is on offline training, in this case by using Deep Deterministic Policy Gradient to generate training data from a simulator (which also outputs the state), this training data then being used for supervised training of the transformer and LSTM state estimator. The authors of [24] have made their code available to use, and we use it as a baseline against which to compare RLT4Rec.

*User Cold Start and RL* A fundamental problem in recommender systems is handling new users. A new user has not yet rated any items and so their preferences are unknown. Traditionally, this has been handled via a special *cold start* recommendation strategy, different from the recommendation strategy used with users who have already rated sufficiently many items. For example, a new user might initially be recommended popular items, items based on user meta-data (such as location, gender) and/or a decision-tree trained from offline data might be used. See [7, 8, 13] for recent surveys on user cold start strategies. Also see [9, 14] for warm-start recommender strategies when users have already rated a few items. In RL recommenders there is no need for a separate cold start strategy. Instead users have different information states and a new user is just treated as having a different state from existing users. If we view the state as embodying the preference of a user, the RL task for a new user is to quickly learn the user's state i.e. to *explore*, and once it is more confident in the state to then make personalised recommendations i.e. to *exploit*, although since there will always be some uncertainty as to the user state the RL system will still continue to explore. The literature on RL for user cold start is, however, quite limited. Probably the most relevant is the recent work by [18, 19] applying Monte Carlo Tree Search (MCTS) to achieve state of the art cold start performance.

## 3 RLT4REC

### 3.1 Learning Task

We consider a recommender system that presents a user with a sequence of items $\{v_i, i = 0, 1, 2, \ldots, t\}$, observing the user rating $r_i$ for item $v_i$ after it has been presented. The rating might be supplied explicitly by the user, e.g. by writing a review, or inferred from e.g. the engagement time they spend viewing a video, listening to a song or reading a news article. The aim of the recommender system is to maximise the sum of the item ratings $R_t = \sum_{i=0}^{t} r_i$. This setup can be extended to incorporate display of a set of items rather than a single item e.g. a slate or full display, and to other choices of utility $R_t$ e.g. to include the cost or value of an item to the recommender itself, to account for the benefit to other users of rating a new item and so on. However, we focus here in the simplest setup with display of single items and sum-rating utility since this already captures most of the important learning aspects and is already quite challenging.

The information state about a user at time $t$ is embodied by the sequence of items that they have viewed and rated up to time $t$ i.e. $\{(v_0, r_0), \ldots, (v_t, r_t)\}$. We typically also have context information such as the approximate location of the user (which country or city), demographics (age, gender) etc. This context information can be readily incorporated either by sharding the RL task, similarly to contextual bandits, or by incorporating the static attributes into the observed sequence of $(v_i, r_i)$ pairs.

*Learning task:* The learning task we consider is therefore as follows: given a sequence of (item, rating) pairs $\{(v_i, r_i), i = 0, \ldots, t\}$ predict the next item $v_{t+1}$ to display so as to maximise the sum-rating $R_t = \sum_{i=0}^{t} r_i$.

*Offline learning:* We have access to training data collected previously. This consists of sequences of (item, rating) pairs but these sequences (i) need not be "good" i.e. they might not have a high sum-rating (e.g. the existing recommender may have poor performance and/or user's may fall back to searching for highly rated items themselves) and (ii) may be "off policy" i.e. are selected according to a different learning strategy from the one that we will learn. We will use this data to initially train the RL system offline, since online training learning purely from live recommendations would likely lead to recommending too many low-rated items during the initial learning phase and so be too irritating for users.

*Unknown user state:* We do not have access to an explicit user state. We do not even know how to characterise this state. All we have access to is the sequence of (item, rating) pairs to date for a user. Note that in most previous offline RL studies, e.g. Decision Transformer [5], the state is provided explicitly by the environment. In RL applied to recommenders previous work has also either assumed availability of an explicit state or has explicitly estimated the state thereby requiring joint design and training of the estimator and reinforcement learner, e.g. see [6, 23, 24, 27, 28]. To the best of our knowledge, RLT4Rec is the first transformer-based RL approach that avoids the need for a state estimate.

## 3.2 RLT4Rec Architecture

We tackle the recommender learning task direcly using a transformer neural network based on a modified GPT architecture [17]. The architecture is illustrated schematically in Figure 1. We refer to this as **R**einforcement **L**earning **T**ransformer for item **Re**commendation (**RLT4Rec**).

This takes as input a sequence of (item, rating) pairs $\{(v_i, r_i), i = 0, \ldots, t-1\}$ plus a target rating $r_t$ for the next item i.e. the last pair in the sequence is $(\cdot, r_t)$ and the task of the neural net is predict the missing item $v_t$ with rating $r_t$. The output of the neural net is a probability vector over the set of possible items. Items predicted to have a higher rating are assigned a higher probability. This is used to make a recommendation by picking the item with highest probability that has not yet been viewed by a user.

*Input Embedding.* We add an embedding layer at the input to map the raw (item, rating) values into dense vectors. Each item token $v_t$ is mapped to a $\mathbb{R}^d$ vector $a_t^{emb}$. Similarly, each rating $r_t$ is mapped to a $\mathbb{R}^d$ vector $r_t^{emb}$. A positional timestep encoding $pe_t$ is then added to these prior to them being fed into the first transformer layer.

*Multi-head Self Attention.* The sequence of input tokens is fed into a stack of $L$ transformer blocks (we use $L = 2$ blocks) with 4 attention heads each. We employ unidirectional transformer blocks with causal attention, which only attend to the previous positions so avoiding any future tokens in the sequences:

$$\text{Attention}(Q, K, V, \text{Mask}) = \text{softmax}\left(\frac{QK^T}{\sqrt{d}} + \text{Mask}\right) V$$

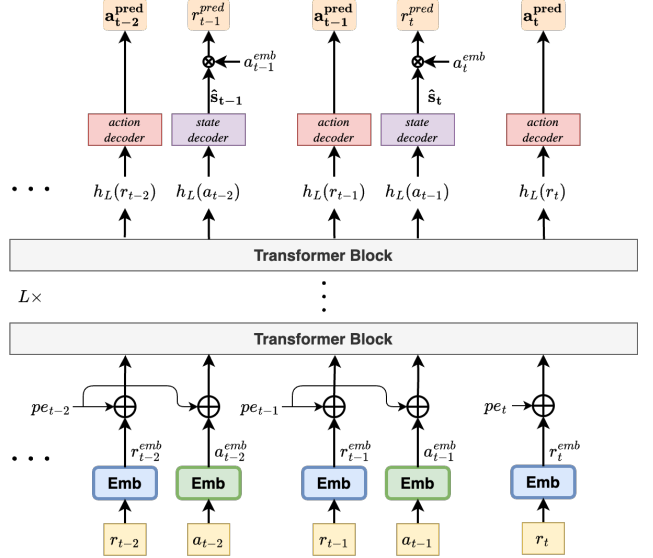

**Figure 1: RLT4Rec architecture**

The output from the multi-head self attention layer is passed through a feed-forward neural layer followed by *GELU* activation layer. We did not find it necessary to uses skip connections.

*Item Prediction.* The output from the last transformer layer is passed through an action decoder module to generate one-step ahead item predictions $\mathbf{a}_i^{\mathbf{pred}}$, $i = 0, \ldots, t$. The action decoder is a linear layer followed by a softmax activation:

$$\mathbf{a}_i^{\mathbf{pred}} = softmax\left(\mathbf{W_a}\, h_L(r_i)\right)$$

where $h_L(r_i)$ is the transformer output corresponding to the input sequence $\{(v_0, r_0), \ldots, (\cdot, r_i)\}$, $\mathbf{a}_i^{\mathbf{pred}}$ is the causal prediction for $v_i$ and is a probability vector over the set of possible items and $W_a$ is a weighting matrix.

To obtain good performance we found it necessary to modify the standard GPT transformer architecture as follows.

*Rating Inner-Product Bottleneck.* During training (see below), the output from the last transformer layer is passed through an inner-product (i.e. matrix factorization) bottleneck to generate one-step ahead rating predictions $\mathbf{r}_i^{\mathbf{pred}}$, $i = 1, \ldots, t$. The inner-product bottleneck consists of a linear layer with a tanh activation followed by an inner-product layer:

$$\hat{\mathbf{s}}_i = tanh\left(\mathbf{W_s}\, h_L(a_{i-1})\right)$$
$$r_i^{pred} = \hat{\mathbf{s}}_i \cdot a_i^{emb}$$

where $h_L(a_{i-1})$ is the transformer output corresponding to the input sequence $\{(v_0, r_0), \ldots, (v_{i-1}, r_{i-1})\}$, $W_s$ is a weighting matrix, $a_i^{emb}$ is the embedding vector for the $i$'th input item and $r_i^{pred}$ is the causal prediction for the rating $r_i$ of the $i$'th item. Following the usual interpretation of matrix factorization we can think of $\hat{\mathbf{s}}_i$ as a user embedding vector.

This bottleneck acts in two ways: (i) it directs the learning of the item embedding $a_i^{emb}$ and (ii) it forces the transformer blocks to learn a vector embedding $\hat{s}_i$ that summarises the observed sequence of (item, rating) pairs. We found the addition of this bottleneck to be essential for obtaining good item predictions. To the best of our knowledge, RLT4Rec is the first transformer-based RL approach that incorporates an inner-product bottleneck.

*Loss Function.* RLT4Rec takes as input a sequence $\{(v_i, r_i), i = 0, \dots, t-1\}$ of (item, rating) pairs plus a target rating $r_t$ for the next item, which during training is taken to be the rating taken from the training data. The output of RLT4Rec is a one-step ahead item prediction vector $\mathbf{a_t^{pred}}$. In addition, during training RLT4Rec also outputs a rating prediction $r_t^{pred}$ (see inner-product bottleneck discussion above). The training loss function used is the sum of (i) the categorical entropy $\mathcal{L}_{CE}(\mathbf{a_t^{pred}}, a_t)$ between the item prediction $\mathbf{a_t^{pred}}$ and the one-hot encoding vector $a_t$ for item $v_t$, and (ii) the square error ($\mathcal{L}_{sqr} = r_t^{pred} - r_t)^2$) between the rating prediction $r_t^{pred}$ and the observed rating $r_t$ for the item. This loss is then averaged over the (item,rating) sequences in the training dataset.

*Training.* RLT4Rec is trained using Adam with learning rate 1e-3, weight decay 1e-5 and gradient clipping (parameter=1.0). For the Netflix, Goodreads, Movielens10M datasets a mini-batch size of 64 was used, while a mini-batch size of 32 was used for the PD1 and PD2 datasets. RLT4Rec was trained for up to 50 epochs and the best validation model was selected for evaluation. We observed that the inner-product bottleneck results in faster convergence as well as better prediction performance. All our training and performance evaluations were conducted on a machine with a NVIDIA GeForce RTX 4090 GPU and AMD Ryzen 5975WX with 32 cpu cores.

An embedding length of $d = 256$ is used for Netflix, Goodreads, Movielens10M and of $d = 128$ for the PD1 and PD2 datasets. We found $L = 2$ transformer blocks with 4 attention heads each to be sufficient to achieve good performance.

# 4 PERFORMANCE EVALUATION

## 4.1 Metrics

*Mean rating across iterations $\overline{R}_{@t}$.* The mean item rating is $\overline{R}_{@t} = \frac{1}{t} \sum_{i=0}^{t-1} r_i$ where $r_i$ is the rating of the $i$'th item presented to a user. This corresponds to the sum-rating utility that we would like to maximise in the RL setup, and so is our primacy performance measure.

*Precision @K.* The precision is the number of relevant items that are within the top-K recommendations. We define an item $v$ to be relevant when the user-rating $r$ of the item is higher than a specified rating threshold, which in our experiments is chosen to be 4.0.

## 4.2 Simulation Environment

To allow us to evaluate recommendation performance in a clean, reproducible manner we use a simulation environment so that we know the ground truth about user ratings.

*User model.* Similarly to [18–20], we use simulation environments derived from measurement datasets consisting of (user, item, ratings) triples. For each dataset we cluster users into groups and estimate the mean $\mu(g, v)$ and variance $\sigma(g, v)^2$ of the ratings by each group $g$ for item $v$. We use the BLC matrix-factorization clustering algorithm [4] for this, although other clustering algorithms (such as k-means) might also be used. To model a user belonging to group $g$, for each item $v$ we draw a random gaussian rating with mean $\mu(g, v)$ and variance $\sigma^2(g, v)$. Note that the user rating for a given item is taken to be fixed since repeated ratings of the same item tend to be highly correlated. In this way we have a full set of item ratings for the user, and can generate as many different users as we like.

*Public datasets.* We use three public datasets, namely the Netflix dataset (480,189 people 17,770 movies, 104M ratings from 1–5), Movielens10M dataset[4] (72K people 10K movies, 10M ratings from 1–5) and the Goodreads10K dataset[5] (53,424 people rating 10,000 books, 5.9M ratings from 1-5).

*Synthetic datasets.* In addition to the above datasets we also use two "toy" synthetic datasets that exhibit extreme behaviour and so allow us to evaluate performance more thoroughly. In the Polarised-1 (PD1) dataset each item tends to be given a high rating by users in just one group, with the users in other groups giving a low rating, e.g. see Table 1(a). The standard deviation of all ratings is $\sigma_g(v) = 0.25$. To achieve good performance (i.e. a high sum-rating) with this dataset a recommender must quickly learn the group that a user belongs to. In the Polarised-2 (PD2) dataset each item tends to be given a high rating by users in multiple groups , e.g. see Table 1(b). The standard deviation of all ratings $\sigma_g(v) = 0.5$. A recommender system can achieve good performance with this dataset without precisely learning the group that a user belongs to.

*Training data.* We create offline training data by generating a population of users (distributed across the different user groups) and then using the simulation environment to generate sequences of (item, user) ratings for these users. The set of items presented to a user are selected uniformly at random. That is, we do not train RLT4Rec on "good" (item, user) sequences yet as we will see below it nevertheless learns to generate good sequences i.e. sequences with high item ratings.

## 4.3 Software

Our RLT4Rec implementation and data is available at https://github.com/dilina-r/RLT4Rec..

## 4.4 Baselines

We evaluate RLT4Rec against four baselines:

(1) *Best\**: This uses a genie that knows the group the user belongs to and so recommends a sequence of items with the highest mean

---

[3]Measurement studies indicate that when people are repeatedly asked to rate the same item on a scale of 1-5 then if they rate 1 or 5 they tend to consistently stick with that rating although when they rate 3 or 4 they may change their rating back and forth between 3 and 4. That is, lumping ratings of 3-4 together in a single bucket these previous studies indicate that a user's rating of an item tends to be consistent.
[4]https://grouplens.org/datasets/movielens/
[5]https://mengtingwan.github.io/data/goodreads

| groups | $v_0$ | ... | $v_{25}$ | ... | $v_{50}$ | ... | $v_{75}$ | ... |
|--------|-------|-----|----------|-----|----------|-----|----------|-----|
| g1 | 5 | | 1 | | 1 | | 1 | |
| g2 | 1 | ... | 5 | ... | 1 | ... | 1 | ... |
| g3 | 1 | | 1 | | 5 | | 1 | |
| g4 | 1 | | 1 | | 1 | | 1 | |

(a) Polarised-1 (PD1) dataset

| group | $v_0$ | ... | $v_{25}$ | ... | $v_{50}$ | ... | $v_{75}$ | ... | $v_{100}$ | ... |
|-------|-------|-----|----------|-----|----------|-----|----------|-----|-----------|-----|
| g1 | 5 | | 1 | | 2 | | 3 | | 4 | |
| g2 | 4 | | 5 | | 1 | | 2 | | 3 | |
| g3 | 3 | ... | 4 | ... | 5 | ... | 1 | ... | 2 | ... |
| g4 | 2 | | 3 | | 4 | | 5 | | 1 | |
| g5 | 1 | | 2 | | 3 | | 4 | | 5 | |

(b) Polarised-2 (PD2) dataset

**Table 1: Example mean ratings for items in the polarised datasets.**

| dataset | target rating | iterations | | | | |
|---------|---------------|------------|------|------|------|------|
| | | 5 | 10 | 15 | 20 | 25 |
| Netflix (8 groups) | 1 | 1.90 | 1.98 | 2.04 | 2.07 | 2.09 |
| | 2 | 2.32 | 2.40 | 2.46 | 2.49 | 2.49 |
| | 3 | 3.16 | 3.18 | 3.21 | 3.24 | 3.24 |
| | 4 | 4.07 | 4.14 | 4.17 | 4.19 | 4.19 |
| | 5 (max) | 4.51 | 4.47 | 4.42 | 4.38 | 4.37 |
| PD2 | 1 | 1.26 | 1.22 | 1.24 | 1.19 | 1.24 |
| | 2 | 2.02 | 2.01 | 2.05 | 2.07 | 2.06 |
| | 3 | 3.00 | 2.96 | 2.94 | 2.90 | 3.06 |
| | 4 | 3.91 | 3.93 | 3.85 | 3.80 | 3.90 |
| | 5 (max) | 4.60 | 4.63 | 4.63 | 4.62 | 4.60 |

**Table 2: Mean rating across $t$ ($\overline{R}_{@t}$) with varying expected returns $\hat{r}_t$ for Netflix (8 groups) and PD2(Polarised-2) Dataset**

ratings for that group, in decreasing order. This is an upper bound on performance, that is not achievable in practice.

(2) *Monte Carlo Tree Search (MCTS)*: This is a strong baseline that represents state of the art performance. It makes use of the user group structure to construct an explicit representation of the information state as a probability vector over the user groups. For a new user all groups are equally likely, but as more user (item, rating) pairs are observed the probability concentrates on the user's group. Given the (item, rating) history to date, the next item to show is chosen by a search of the lookahead tree. The lookahead tree grows exponentially and so MCTS uses a UCB-like approach to focus on exploring the most promising branches. It uses knowledge of the mean and variance of the item ratings in each group to estimate the reward of each branch. See [18, 19] for further details. In summary, MCTS has the advantage of extra knowledge of the RL task (the number of user groups and the mean and variance of the item ratings in each group) that is not available to RLT4Rec, and uses this to directly search for solutions to the RL task. When run for a sufficient number of iterations it is guaranteed to find the optimal solution.

(3) *Random-Uniform (R-U)*: This presents items selected uniformly at random from the items not yet viewed by the user.

(4) *Random-Popular (R-P)*: This selects items with probability proportional to the number of times they have been rated by users. Popular items tend to have higher ratings.

(5) *Decision Tree (D-Tree)*: CART based Decision Tree, optimised to minimise Regret in user-cold start learning.

(6) *Prompt-based Reinforcement Learning [24] (PRL)*: An offline-RL approach for next-item recommendation. In our experiments, we use the GRU variant of the PRL recommender.

(7) *Decision Transformer (D-Tf)*: This is a vanilla Decision Transformer [5], employed with a GRU-network to estimate the state at each timestep.

### 4.5 Impact of varying the target item rating

RLT4Rec takes as input a sequence $\{(v_i, r_i), i = 0, \ldots, t-1\}$ of (item, rating) pairs plus a target rating $r_t$ for the next item. It outputs a one-step ahead prediction of the next item $v_t$ with rating $r_t$. As a sanity check on performance, Table 2 shows measurements of the mean rating of the predicted item $v_t$ as the target rating $r_t$ is varied. Data is shown for the Netflix and PD2 datasets but similar results are also observed for the other datasets.

For these two datasets ratings lie in the range 1 to 5. It can be seen that the mean rating of $v_t$ quite closely tracks the target rating $r_t$ over this range, except for the Netflix dataset when the target rating is 1 in which case the mean rating of $v_t$ is about 2. The is because the Netflix dataset contains very few items with rating 1 (the ratings tend to be bunched between 3 and 5).

In summary, the mean rating of the one-step ahead item prediction by RLT4Rec closely matches the target rating $r_t$ input to RLT4Rec. For the rest of this paper we set the target rating to be the maximum rating for each dataset, e.g. $r_t = 5$ for Netflix.

### 4.6 Recommending for cold-start users

When a new user joins the system it initially has no knowledge of the preferences of the user and so would like to quickly learn these. Traditionally, the recommender system therefore initially starts in an "exploration" phase where the first few items that it asks the new user to rate are chosen with the aim of discovering the user's preferences. For example, one common approach to this new user cold-start task is to take ratings already collected from a population of users, use these to cluster users into groups and then train a decision-tree to learn a mapping from item ratings to the user group. When a new user joins the system this decision-tree is used to decide which items the user is initially asked to rate and in this way the group to which the user belongs is initially estimated. The recommender system then switches to "normal" operation to make subsequent recommendations.

RLT4Rec treats cold start and normal operation in an integrated manner. In all cases the input to RLT4Rec is a sequence of user (item, rating) pairs that it uses to recommend the next item. For a new user the sequence is empty, and then grows over time as the user rates more items. *There is no need for a separate cold start mode with RLT4Rec.*

Since RLT4Rec seeks to make good item predictions for the user, when the input (item, rating) sequence is short RLT4Rec automatically tends to "explore" by choosing items that quickly reveal the user's preferences and allow RLT4Rec to learn them. This can be seen in Table 3 and Figure 2, which shows the mean rating of the recommended item versus the number of items the user has rated (averaged over 200 users per group - the results are similar when we also use more users). Also shown in Table 3 and Figure 2 are the results for Decision Tree, and MCTS which achieves state of the art performance for this cold-start task [18, 19] and so provides an extremely strong baseline. It can be seen that the RLT4Rec outperforms all baselines, including the MCTS, yielding higher rewards on new users even on pure-start conditions.

This good cold start performance is achieved despite the fact that MCTS has the advantage of being provided with knowledge of the user groups, including the item rating means and variances for each user group, whereas RLT4Rec lacks any information about the user groups and must instead learn it from the training data.

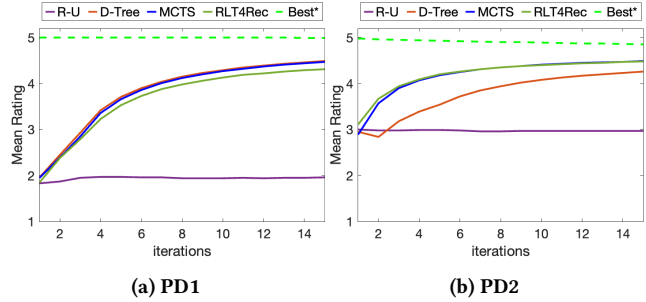

**(a) PD1**                **(b) PD2**

**Figure 2: Mean rating across $t$ ($\overline{R}_{@t}$) vs number of items $t$ - Polarised Datasets PD1 and PD2.**

Looking at the data in Figure 2 in more detail, recall that in the PD1 dataset each item is rated highly by just one user group, and given low ratings by users in the other groups. To achieve good performance it is therefore essential for the recommender to quickly and accurately learn which group a user belongs to. It can be seen that Decision Tree and MCTS learns to recommend highly rated items after about 5 items (after that the mean rating is always close to the maximum rating of 5). A similar behaviour can be seen with the RLT4Rec. In contrast, picking random items never achieves a mean rating of more than about 2. In the PD2 dataset each item is given a high rating by multiple user groups, and so good performance can be achieved without accurately learning the group that a user belongs to. Reflecting this, it can be seen in Figure 2 that random items now achieve a mean rating of 3 and both RLT4Rec and MCTS learn to recommend highly rated items after only about 2 items.

| dataset | algo | iterations | | | | |
|---|---|---|---|---|---|---|
| | | *5* | *10* | *15* | *20* | *25* |
| Netflix (8 groups) | Best* | 5.00 | 5.00 | 5.00 | 5.00 | 4.99 |
| | R-U | 3.19 | 3.20 | 3.21 | 3.21 | 3.21 |
| | R-P | 4.16 | 4.15 | 4.14 | 4.13 | 4.12 |
| | D-Tree | 3.33 | 3.59 | 3.87 | 4.00 | 4.07 |
| | MCTS | 3.84 | 3.97 | 4.02 | 4.05 | 4.07 |
| | RLT4Rec | **4.51** | **4.47** | **4.42** | **4.38** | **4.37** |
| Goodreads8 (8 groups) | Best* | 5.00 | 4.99 | 4.97 | 4.95 | 4.94 |
| | R-U | 3.58 | 3.57 | 3.56 | 3.55 | 3.55 |
| | R-P | 3.63 | 3.63 | 3.63 | 3.63 | 3.63 |
| | D-Tree | 3.44 | 3.68 | 3.84 | 3.91 | 3.94 |
| | MCTS | 3.84 | 3.88 | 3.87 | 3.86 | 3.85 |
| | RLT4Rec | **4.21** | **4.20** | **4.17** | **4.14** | **4.12** |
| Movielens (8 groups) | Best* | 5.00 | 5.00 | 5.00 | 4.99 | 4.98 |
| | R-U | 3.16 | 3.16 | 3.15 | 3.16 | 3.16 |
| | R-P | 3.60 | 3.59 | 3.59 | 3.59 | 3.58 |
| | D-Tree | 3.04 | 3.39 | 3.66 | 3.78 | 3.85 |
| | MCTS | 3.77 | 3.83 | 3.86 | 3.87 | 3.89 |
| | RLT4Rec | **4.14** | **4.12** | **4.11** | **4.10** | **4.09** |
| Netflix (16 groups) | Best* | 5.00 | 5.00 | 5.00 | 5.00 | 5.00 |
| | R-U | 3.29 | 3.30 | 3.30 | 3.30 | 3.30 |
| | R-P | 3.73 | 3.71 | 3.72 | 3.72 | 3.71 |
| | D-Tree | 3.21 | 3.49 | 3.80 | 3.94 | 4.02 |
| | MCTS | 3.84 | 3.93 | 3.97 | 4.00 | 4.02 |
| | RLT4Rec | **4.33** | **4.35** | **4.34** | **4.32** | **4.30** |
| Goodreads (16 groups) | Best* | 4.98 | 4.95 | 4.92 | 4.88 | 4.85 |
| | R-U | 3.58 | 3.58 | 3.58 | 3.58 | 3.58 |
| | R-P | 3.72 | 3.72 | 3.72 | 3.72 | 3.71 |
| | D-Tree | 3.52 | 3.66 | 3.81 | 3.87 | 3.89 |
| | MCTS | 3.82 | 3.86 | 3.88 | 3.89 | 3.89 |
| | RLT4Rec | **4.05** | **4.07** | **4.05** | **4.02** | **3.99** |
| Movielens10M (16 groups) | Best* | 5.00 | 5.00 | 4.99 | 4.98 | 4.97 |
| | R-U | 3.36 | 3.36 | 3.36 | 3.36 | 3.36 |
| | R-P | 3.53 | 3.53 | 3.53 | 3.53 | 3.53 |
| | D-Tree | 3.65 | 3.57 | 3.67 | 3.79 | 3.86 |
| | MCTS | 3.85 | 3.89 | 3.92 | 3.94 | 3.95 |
| | RLT4Rec | **4.12** | **4.10** | **4.07** | **4.05** | **4.01** |

**Table 3: Mean ratings across $t$ recommendations ($\overline{R}_{@t}$) with Best\*, R-U(Random-Uniform), R-P(Random-Popular), D-Tree, MCTS and RLT4Rec for Netflix, Goodreads, Movielens datasets with 8 and 16 groups**

### 4.7 Performance for established users

Table 4 and Figure 3 show measurements of the longer-term performance of RLT4Rec for the Netflix, Goodreads, Movielens datasets (averaged over 200 users per group). In Table 3 the recommender that achieves the highest mean item rating is highlighted in bold (excluding the *Best\** upper bound, which is unachievable). It can be seen from Table 3 that the performance of RLT4Rec consistently dominates that of MCTS and Decision Tree. This behaviour is also evident in Figure 3, which extends the data out to 50 items and also compares the performance against the R-P and R-U strategy.

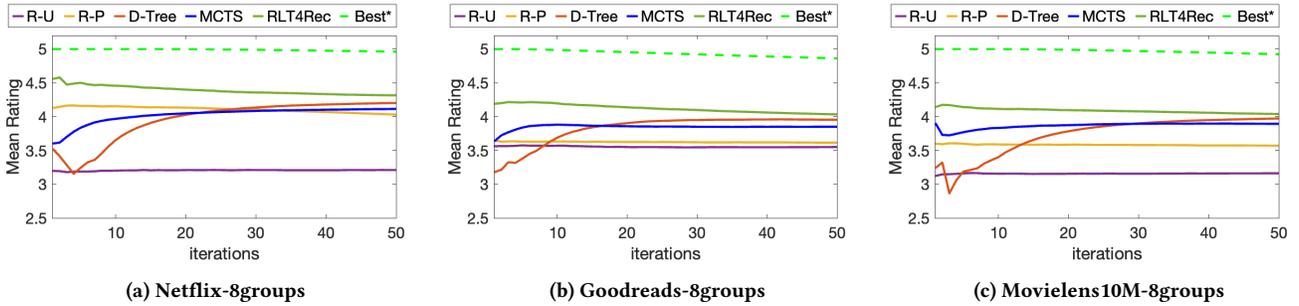

**Figure 3: Mean rating across $t$ ($\overline{R}_{@t}$) up to 50 items, for Netflix, Goodreads, Movilens10M with 8 groups**

We also present measurements of the list recommendation performance of RLT4Rec after a user has rated 25 or more items (we take this as a rough threshold whereby a user is no longer "new"). A list of recommended items is generated by predicting a relevance score for all the items not yet viewed by the user and recommending the top $K$ items. Recall that RLT4Rec outputs a probability vector over the set of possible items, assigning higher probability to items predicted to be given a higher rating by the user. We use this probability as the relevance score for RLT4Rec. For the MCTS, we use the estimated mean reward, computed for each item during the tree search, as the relevance score. We also compare the performance against two offline-RL methods PRL [24] and a vanilla Decision Transformer [5].

Table 4 shows the measured Precision@$K$ performance as the size $K$ of the list of recommended item is varied between 1 and 20 ($K$=1 represents the case when only a single item is recommended). It can be seen that again RLT4Rec consistently dominates that of PRL, Decision Transformer, MCTS, R-U and R-P.

The Decision Transformer is provided with the same information as our RLT4Rec in the input sequence, along with an additional 'state' which is generated through a GRU network. It can be seen that the decision transformer outperforms the PRL method, which also uses a GRU for state-estimation and predicts an item for a given expected reward. However, RLT4Rec shows a substantial improvement over all methods, despite not being given a state estimate in the input sequence. That is, RLT4Rec is both simpler and more effective.

It is worth noting that the performance gain of RLT4Rec over MCTS is achieved despite the fact that MCTS has the advantage over RLT4Rec of knowledge of the RL task (the number of user groups and the mean and variance of the item ratings in each group) that is not available to RLT4Rec, and uses this to directly search for solutions to the RL task. When run for a sufficient number of iterations it is guaranteed to find the optimal solution. However, the computational cost of MCTS increases exponentially with the size of the set of possible items to recommend (since the lookahead search tree grows exponentially), and acts to constrain the performance achievable in practice.

| dataset | algo | prec@1 | prec@5 | prec@10 | prec@15 | prec@20 |
|---|---|---|---|---|---|---|
| Netflix (8 groups) | R-U | 0.346 | 0.342 | 0.340 | 0.340 | 0.341 |
| | R-P | 0.663 | 0.675 | 0.675 | 0.667 | 0.661 |
| | MCTS | 0.653 | 0.660 | 0.642 | 0.640 | 0.639 |
| | PRL-GRU | 0.683 | 0.763 | 0.759 | 0.753 | 0.751 |
| | D-Tf | 0.826 | 0.800 | 0.776 | 0.758 | 0.750 |
| | RLT4Rec | **0.855** | **0.830** | **0.794** | **0.770** | **0.758** |
| Goodreads (8 groups) | R-U | 0.393 | 0.412 | 0.414 | 0.413 | 0.414 |
| | R-P | 0.423 | 0.442 | 0.441 | 0.438 | 0.439 |
| | MCTS | 0.561 | 0.551 | 0.537 | 0.527 | 0.521 |
| | PRL-GRU | 0.696 | 0.641 | 0.624 | 0.616 | 0.612 |
| | D-Tf | 0.699 | 0.659 | 0.639 | 0.628 | 0.617 |
| | RLT4Rec | **0.778** | **0.718** | **0.696** | **0.675** | **0.660** |
| Movielens10M (8 groups) | R-U | 0.267 | 0.267 | 0.270 | 0.270 | 0.268 |
| | R-P | 0.393 | 0.392 | 0.392 | 0.392 | 0.391 |
| | MCTS | 0.581 | 0.531 | 0.525 | 0.517 | 0.516 |
| | PRL-GRU | 0.618 | 0.598 | 0.585 | 0.572 | 0.562 |
| | D-Tf | 0.648 | 0.627 | 0.612 | 0.602 | 0.594 |
| | RLT4Rec | **0.693** | **0.653** | **0.632** | **0.622** | **0.615** |

**Table 4: Precision@K for list of top-K recommendations for R-U(Random-Uniform), R-P(Random-Popular), MCTS, PRL-GRU, D-Tf(Decision Transformer) and our RLT4Rec with established users in Netflix, Goodreads, Movilen10M datasets (8 user-groups)**

## 5 RLT4REC LEARNS AN INFORMATION STATE REPRESENTATION

In contrast to conventional RL systems (including previous approaches using transformers), the RL state is not provided as an input to RLT4Rec. All that is provided as input to RLT4Rec is a sequence of user (item,rating) pairs. Nevertheless, RLT4Rec succeeds at accurately predicting highly rated items, even for polarised datasets such as PD1 where each item is rated highly by just one user group and so for good performance it is essential that the recommender learns which group a user belongs to. It therefore seems that RLT4Rec learns an internal representation of the user information state that it then uses when making predictions. In this section we investigate this further.

### 5.1 User Information State

For the RL task that we consider here, the relevant state is the user information state. In particular, in the user model in Section 4 each user belongs to one of a set of groups and the information state after rating a sequence of $t$ items is the vector $P^{(t)} := (p_1^{(t)}, \ldots, p_n^{(t)})$ where $p_g^{(t)}$ is the probability that the user belongs to group $g$ given the observed item ratings.

We can calculate the ground-truth value of this state vector as follows. The rating of item $v$ by a user in group $g$ is a Gaussian random variable with mean $\mu(g, v)$ and variance $\sigma^2(g, v)$. Denoting the user's group by random variable $G$ and the user's rating of item $v$ by random variable $R(v)$, we therefore have that

$$p(R(v) = r | G = g) = (1/\sqrt{2\pi}\sigma(g, v))e^{-(r-\mu(g,v))^2/2\sigma^2(g,v)}$$

and for observed sequence $D^{(t)}$ of ratings $r(v_1), r(v_2), \ldots, r(v_t)$ for items $\mathcal{V}^{(t)} = \{v_1, v_2, \ldots, v_t\}$ it follows that $p(D^{(t)} | G = g) = \gamma^{(t)}(g)e^{-L^{(t)}(g)}$ where $L^{(t)}(g) := \sum_{i=1}^{t}(r(v_i) - \mu(g, v_i))^2/2\sigma^2(g, v_i)$, $\gamma^{(t)}(g)) := 1/(2\pi)^{t/2} \times 1/\Pi_{i=1}^{t}\sigma(g, v_i)$. By Bayes rule,

$$p_g^{(t)} = \frac{p(D^{(t)} | G = g)p(G = g)}{\sum_{h \in \mathcal{G}} p(D^{(t)} | G = h)p(G = h)} \quad (1)$$

with $p(D^{(t)}) = \sum_{h \in \mathcal{G}} p(D^{(t)} | G = h)p(G = h)$. Assuming uniform prior $p_g^{(0)} = p(G = g) = 1/|\mathcal{G}|$ then $p_g^{(t)} = \frac{\gamma^{(t)}(g)e^{-L^{(t)}(g)}}{\sum_{h \in \mathcal{G}} \gamma^{(t)}(h)e^{-L^{(t)}(h)}}$ for $t = 1, 2, \ldots$.

## 5.2 Reconstructing the State From RLT4Rec Activations

To investigate whether RLT4Rec learns the state $P^{(t)}$ corresponding to an (item, rating) sequence we use a *probe* [2, 3]. Namely, we train a 2 layer MLP to predict the state $P^{(t)}$ from the neuron activations within RLT4Rec. In particular, we use the $\hat{s}_t$ activations from the inner-product bottleneck in RLT4Rec as input to the probe MLP, since these are the obvious analogous to a user embedding vector within RLT4Rec. We create training data for the probe by generating (item, rating) sequences for random items and users in each group.

If the probe MLP is able to accurately predict the state $P^{(t)}$ from the RLT4Rec internal activations, this suggests that RLT4Rec has indeed learned a representation of the state. As a baseline for comparison we extract the hidden layer outputs from an untrained RLT4Rec model, initialised with random model parameters.

## 5.3 State Reconstruction Results

Table 5 shows the measured Mean Absolute Error $\frac{1}{|\mathcal{G}|}\sum_{g \in \mathcal{G}} \left| P_g^{(t)} - \hat{P}_g^{(t)} \right|$ between the group membership $\hat{P}^{(t)}$ predicted by the probe MLP and the true group membership probabilities $P^{(t)}$. It can be seen that after training the MAE for RLT4Rec is generally an order of magnitude smaller than before training, consistent with the hypothesis that trained RLT4Rec model is indeed representing the state within its internal $\hat{s}_t$ activations.

Table 6 also shows the measured group estimation accuracy i.e. the fraction of times that the user group $\hat{g} = \arg\max_{h \in \mathcal{G}} P_h^{(t)}$ estimated to have highest probability by the probe is in fact the correct user group. It can be seen that after training, the accuracy of RLT4Rec rise to be close to 100% after 5 items have been rated whereas for the untrained RLT4Rec the accuracy remains around 25% regardless of the number of items rated. Again, this evidence is consistent with the hypothesis that trained RLT4Rec model has learned the user state.

| dataset | method | iterations | | | | |
|---|---|---|---|---|---|---|
| | | 1 | 2 | 3 | 4 | 5 |
| PD1 | Random | 0.199 | 0.276 | 0.321 | 0.345 | 0.359 |
| | Trained RLT4Rec | 0.003 | 0.003 | 0.005 | 0.005 | 0.005 |
| PD2 | Random | 0.231 | 0.274 | 0.293 | 0.301 | 0.305 |
| | Trained RLT4Rec | 0.053 | 0.035 | 0.022 | 0.016 | 0.012 |

**Table 5: Mean Absolute Error for Probe $\hat{P}_G^{(t)}$ predictions - PD1 and PD2 Datasets**

| dataset | method | iterations | | | | |
|---|---|---|---|---|---|---|
| | | 1 | 2 | 3 | 4 | 5 |
| PD1 | Probe-Random | 0.250 | 0.262 | 0.256 | 0.252 | 0.254 |
| | Probe-RLT4Rec | 0.506 | 0.676 | 0.826 | 0.907 | 0.954 |
| | True $P_g^{(t)}$ | 0.502 | 0.682 | 0.822 | 0.909 | 0.957 |
| PD2 | Probe-Random | 0.239 | 0.248 | 0.256 | 0.264 | 0.271 |
| | Probe-RLT4Rec | 0.729 | 0.875 | 0.940 | 0.967 | 0.982 |
| | True $P_g^{(t)}$ | 0.748 | 0.889 | 0.949 | 0.972 | 0.988 |

**Table 6: Group-estimation accuracy, calculated from predicted Group Membership $\hat{P}_g^{(t)}$ with Random (Untrained) and RLT4Rec (trained) against the True Probabilities $P_g^{(t)}$, for PD1 and PD2 Datasets**

## 6 CONCLUSION

In summary, we introduce a new sequential transformer architecture RLT4Rec and demonstrate that it achieves state of the art performance in a range of item recommendation tasks.

RLT4Rec uses a relatively simple transformer architecture that takes as input a sequence of user (item, rating) pairs and outputs the next item to present to the user. The RLT4Rec architecture includes an inner-product bottleneck, and we find that this plays a key role in achieving good performance. In particular, our analysis indicates that this bottleneck allows RLT4Rec to construct an internal state representation that successfully captures user preferences. There is no need to input an explicit state observation or estimate to RLT4Rec. This not only simplifies the recommender system architecture but also avoids the difficult design step of deciding upon an appropriate user state representation since this is now learned automatically.

We show that RLT4Rec handles new users and established users within the same consistent framework and automatically balances the "exploration" needed to discover the preferences of a new user with the "exploitation" that is more appropriate for established users. RLT4Rec consistently improves upon the performance of PRL [24] for established users and MCTS [19] for new users.

Training of RLT4Rec is robust and fast and is insensitive to the choice of training data. It can be trained using random sequences of items yet nevertheless learns to generate "good" personalised sequences that the user tends to rate highly. While offline RL learning is known to be difficult, our results show that this may be much less troublesome for RLT4Rec.

## ACKNOWLEDGEMENT

This work was supported by Science Foundation Ireland (SFI) under grant 16/IA/4610